\documentclass[11pt]{article}
\newcommand{\Comment}[1]{{}}
\usepackage{amsfonts,amsthm,amsmath,amssymb,slashed}
\usepackage[textwidth = 430 pt, textheight = 630 pt]{geometry}

\Comment{\usepackage{color}
\definecolor{MyDarkBlue}{rgb}{0.15,0.15,0.45}
\usepackage[linktocpage=true]{hyperref}
\hypersetup{
colorlinks=true,
citecolor=MyDarkBlue,
linkcolor=MyDarkBlue,
urlcolor=MyDarkBlue,
pdfauthor={Horatiu Nastase and Howard J. Schnitzer},
pdftitle={On KLT and SYM-supergravity relations from 5-point 1-loop amplitudes},
pdfsubject={hep-th}
}

\usepackage[numbers,sort&compress]{natbib}
\usepackage{hypernat}}

\newcommand\ignore[1]{}
\def\one{{\,\hbox{1\kern-.8mm l}}}

\def\Tr{{\rm Tr\, }}

\def\a{\alpha}\def\b{\beta}

\newcommand{\Cset}{{\,\,{{{^{_{\pmb{\mid}}}}\kern-.45em{\mathrm C}}}}}

\newcommand{\be}{\begin{equation}}
\newcommand{\bea}{\begin{eqnarray}}

\newcommand{\ee}{\end{equation}}
\newcommand{\eea}{\end{eqnarray}}

\begin{document}

\renewcommand{\thefootnote}{\fnsymbol{footnote}}

\makeatletter
\@addtoreset{equation}{section}
\makeatother
\renewcommand{\theequation}{\thesection.\arabic{equation}}

\rightline{}
\rightline{}
   \vspace{1.8truecm}

\begin{flushright}
BRX-TH-625
\end{flushright}

\vspace{10pt}

%%%%%%%%%%%%%%%%%

\begin{center}
{\LARGE \bf{\sc On KLT and SYM-supergravity relations \\
from 5-point 1-loop amplitudes}}
\end{center} 
 \vspace{1truecm}
\thispagestyle{empty} \centerline{
    {\large \bf {\sc Horatiu Nastase${}^{a,}$}}\footnote{E-mail address: \Comment{\href{mailto:nastase@ift.unesp.br}}{\tt 
    nastase@ift.unesp.br}} {\bf{\sc and}}
    {\large \bf {\sc Howard J. Schnitzer${}^{b,}$}}\footnote{E-mail address:
                                \Comment{ \href{mailto:schnitzr@brandeis.edu}}{\tt schnitzr@brandeis.edu}}
                                                           }

\vspace{1cm}
\centerline{{\it ${}^a$ 
Instituto de F\'{i}sica Te\'{o}rica, UNESP-Universidade Estadual Paulista}} \centerline{{\it 
R. Dr. Bento T. Ferraz 271, Bl. II, Sao Paulo 01140-070, SP, Brazil}}

\vspace{.8cm}
\centerline{{\it ${}^b$ 
Theoretical Physics Group, Martin Fisher School of Physics}} \centerline{{\it Brandeis University, Waltham, 
MA 02454, USA}}

\vspace{2truecm}

%%%%%%%%%%%%%%%%%
\thispagestyle{empty}

\centerline{\sc Abstract}

\vspace{.4truecm}

\begin{center}
\begin{minipage}[c]{380pt}{\noindent We derive a new non-singular tree-level KLT relation for the $n=5$-point amplitudes, with manifest 
$2(n-2)!$ symmetry, using information from one-loop amplitudes and IR divergences, and speculate how one might extend it to higher n-point
functions. 
We show that the subleading-color ${\cal N}=4$ SYM 5-point amplitude has leading IR divergence of $1/\epsilon$, which is essential for the 
applications of this paper. We also propose a relation
between the subleading-color ${\cal N}=4$ SYM and ${\cal N}=8$ supergravity 1-loop 5-point amplitudes, valid for the IR divergences and possibly 
for the whole amplitudes, using techniques similar to those used in our derivation of the new KLT relation.
}
\end{minipage}
\end{center}

\vspace{.5cm}

\setcounter{page}{0}
\setcounter{tocdepth}{2}

\newpage

%\tableofcontents
\renewcommand{\thefootnote}{\arabic{footnote}}
\setcounter{footnote}{0}

\linespread{1.1}
\parskip 4pt

{}~
{}~

\section{Introduction}

In the last several years there have been enormous advances in the methods for dealing with the large $N$ (i.e. planar) sector of 
${\cal N}=4$ $SU(N)$ SYM theory. There also has been considerable progress in the understanding of ${\cal N}=8$ supergravity, in part 
because of the surprisingly related structure of the two theories. One of the pioneering connections between SYM and supergravity theories 
are the KLT relations \cite{Kawai:1985xq}, originally proved using string theory methods \cite{Kawai:1985xq,Berends:1988zp}. 

More recently, alternate versions of KLT relations have been presented based on field theoretic techniques at the tree level
\cite{BjerrumBohr:2010ta,BjerrumBohr:2010yc}. One form of 
these new relations has manifest $(n-3)!$ permutation symmetry for the $n$-point functions, and another has $(n-2)!$ symmetry, but requires 
regularization as a consequence of singularities. These are reviewed in section 2. They are part of a flurry of recent activity
relating ${\cal N}=4$ SYM and ${\cal N}=8$ supergravity, including 
\cite{Elvang:2007sg,Ananth:2007zy,Feng:2010br,Bern:2010ue,BjerrumBohr:2010hn,Bern:2008qj,Vaman:2010ez,Bern:2010yg}
(among older works see also \cite{Bern:1998sv,Bern:1994zx,Bern:1996ja}).
Recent work applying the KLT relations include \cite{BjerrumBohr:2010zb,Feng:2010hd,Tye:2010kg,Elvang:2010kc}.
In section 3 we review the 1-loop expansion of the 5-point 
${\cal N}=8$ supergravity amplitudes, and the single and double trace ${\cal N}=4$ SYM amplitudes; noting that $A_{5;3}$ and the 1-loop 
supergravity amplitude both have $1/\epsilon$ IR divergences.

In section 4 we present a new tree-level KLT relation for the $n=5$-point amplitudes, using information from one-loop SYM and supergravity 
amplitudes and their IR divergences. This results in a KLT relation for 5-point functions with $2(n-2)!$ manifest symmetry, without the need 
for regularization. Our KLT relations are proved explicitly in section 5 using the helicity spinor formalism and the Parke-Taylor formula. 
In section 6, in analogy with our previous work \cite{Naculich:2008ys,Naculich:2008ew} 
on 4-point functions of ${\cal N}=8$ supergravity and subleading color ${\cal N}=4$ SYM theories, 
both with the $1/\epsilon$ IR divergence, we explore the possibility that the 1-loop 5-point supergravity amplitude can be expressed as a linear 
combination of the $A_{5;3}$ SYM amplitudes. In particular we propose a linear relation among the $1/\epsilon$ IR divergences of the two theories. 
In section 7 we present suggestions for future work and our conclusions in section 8.

%In this paper we will use 1-loop information to find new relations between the amplitudes of ${\cal N}=4$ SYM 
%and ${\cal N}=8$ supergravity.

\section{KLT relations and amplitude basis}\label{kltsection}

In this section, we quickly review what is known about SYM amplitudes and the KLT relations at tree level.

At tree  level, there are quadratic relations between the $n$-point amplitudes of ${\cal N}=4$ SYM and those of 
${\cal N}=8$ supergravity, known as KLT relations. In these relations, the helicity information is all contained within the amplitudes, and the 
coefficients are all function of the kinematic invariants $s_{ij}$ only.

The color-dressed tree amplitude ${\cal A}_n$ of ${\cal N}=4$ SYM is related to the color-ordered tree amplitude $A_n$ 
by (using the notation in \cite{Bern:2008qj})
\bea
{\cal A}_n^{tree}(12...n)&=&g^{n-2}\sum_{\sigma\in S_n/Z_n}\Tr(T^{a_{\sigma(1)}}...T^{a_{\sigma(n)}})A_n^{tree}(\sigma(1)...\sigma(n))\cr
&=&g^{n-2}\sum_{P(23...n)}\Tr[T^{a_1}T^{a_{P(2)}}...T^{a_{P(n)}}]A_n^{tree}(12...n)
\eea
where $1$ is fixed and $P(23..n)$ is a permutation of $2,3,...,n$. We define
\be
\tilde f ^{abc}=i\sqrt{2}f^{abc}=\Tr([T^a,T^b]T^c)
\ee
Note that from now on, we will drop the index $tree$ from the SYM and supergravity amplitudes, and keep the $1-loop$ index on the 1-loop amplitudes 
only. 

For the $n$-point color-ordered tree amplitudes, there is a basis of $(n-2)!$ amplitudes out of the total $n!$, 
called Kleis-Kuijf basis, and we can 
find the others easily in terms of it \cite{Bern:2008qj}. It is based on the existence of the Kleis-Kuijf relations (KK), which are 
\be
A_n(1,\{\a\},n,\{\b\})=(-1)^{n_\b}\sum_{\{\sigma\}_i\in OP(\{\a\},\{\b^T\})}A_n(1,\{\sigma\}_i,n)\label{KK}
\ee
where $\sigma_i$ are ordered permutations, i.e. that keep the order of $\{\a\}$ and of $\{\b^T\}$ inside $\sigma_i$. 
Thus the KK basis is $A_n(1,{\cal P}(2,...,n-1),n)$ where ${\cal P}$ are arbitrary permutations.

Using cyclicity, reflection invariance,
\be
A_n(12...n)=(-1)^nA_n(n...21)
\ee
and the KK relations (\ref{KK}), we can reconstruct all the $n!$ amplitudes from the KK basis.

The color-ordered amplitudes $A_n(k)$ of the KK basis are written as sums over Feynman diagrams, each 
giving a numerator and a multiple pole, of the generic type
\be
A_n(k)=\sum_{a_k} \frac{n_{a_k}}{(\prod s_l)_{a_k}}
\ee

Then the color-dressed amplitude is
\be
{\cal A}_n=\sum_i\frac{sgn_ic_i n_i}{(\prod_j s_j)_i}
\ee
where $sgn_i$ are signs, $c_i$ are color factors, made up of Feynman diagrams for $\tilde f^{abc}$, 
satisfying Jacobi-type identities,
\be
c_i+c_j+c_k=0
\ee
One can then find, as Bern, Carrasco and Johansson (BCJ) showed \cite{Bern:2008qj}, 
corresponding numerators satisfying satisfying the same Jacobi-type identities,
\be
n_i+n_j+n_k=0
\ee
which can be solved in terms of $(n-2)!$ independent numerators, leading to a linear relation in the KK basis
$|A>=M|N>$. Then one obtains graviton scattering amplitudes as 
\be
M_{n,sugra}=\sum_i \frac{n_i \tilde n_i}{(\prod_j s_j)_i}
\ee
where $(\tilde c_i,\tilde n_i)$ correspond to another or the same gauge theory. 
There are generalized gauge invariances acting on the numerators which leave the Jacobi identities intact, such that 
one can find $(n-3)!$ independent numerators or amplitude from which to generate the rest. 

The KLT relations relate graviton tree amplitudes with sums of squares (products) of gauge tree amplitudes. 
The original KLT relations were derived from string theory in the $\a'\rightarrow 0$ limit \cite{Kawai:1985xq,Berends:1988zp}, and 
can be expressed as (we use the notation of \cite{Bern:1998sv})
\bea
M_n(12...n)&=&(-1)^{n+1}\Big[A_n(12...n)\sum_{perms}f(i_1...i_j)\bar f(l_1...l_{j'})\times\cr
&&\times A_n(i_1,...,i_j,1,n-1,l_1,...,l_{j'},n)\Big]+{\cal P}(2,...,n-2)\cr
f(i_1,...,i_j)&=&s(1,i_j)\prod_{m=1}^{j-1}\Big(s(1,i_m)+\sum_{k=m+1}^jg(i_m,i_k)\Big)\cr
\bar f(l_1,...,l_{j'})&=&s(l_1,n-1)\prod_{m=2}^{j'}\Big(s(l_m,n-1)+\sum_{k=1}^{m-1}g(l_k,l_m)\Big)
\eea
where 'perms' are $(i_1,...,i_j)\in {\cal P}(2,...,n/2)$, $(l_1,...,l_{j'}\in{\cal P}(n/2+1,...,n-2)$, $j=n/2-1,j'=n/2-2$ and 
$g_{i,j}=s_{ij}$ if $i>j$ and zero otherwise.

In \cite{BjerrumBohr:2010ta} and \cite{BjerrumBohr:2010yc}, new forms of the KLT relations for any $n$-point function were found. 
They are both written in terms of the functions 
\bea
&&{\cal S}[i_1...i_k|j_1...j_k]=\prod_{t=1}^k(s_{i_t1}+\sum_{q>t}^k\theta(i_t,i_q)s_{i_ti_q})\cr
&&\tilde {\cal S}[i_1...i_k|j_1...j_k]=\prod_{t=1}^k(s_{j_tn}+\sum_{q<t}^k\theta(j_q,j_t)s_{j_qj_t})
\eea
where $\theta(i_t,i_q)$ is zero in $(i_t,i_q)$ has the same order in both sets ${\cal I}=\{i_1,...,i_k\}$ and 
${\cal J}=\{j_1,...,j_k\}$ and is 1 otherwise, and similarly for $\theta(j_q,j_t)$. 

A form of KLT relations was found in \cite{BjerrumBohr:2010ta}, but need to be regularized,
due to a singular denominator,
\bea
M_n&=&(-1)^n \sum_{\gamma, \b}\frac{ {\tilde A}_n(n,\gamma_{2,n-1},1){\cal S}[\gamma_{2,n-1},\beta_{2,n-1}]_{p_1}
A_n(1,\beta_{2,n-1},n)}{s_{12...n-1}}\cr
M_n&=&(-1)^n\sum_{\b, \gamma}\frac{  A_n(n,\b_{2,n-1},1)\tilde {\cal S}[\b_{2,n-1},\gamma_{2,n-1}]_{p_n}
\tilde A_n(1,\gamma_{2,n-1},n)}{s_{23...n}}\label{KLTold}
\eea
however they have a large $(n-2)!$ manifest symmetry. Another set was proven in \cite{BjerrumBohr:2010yc} which is non-singular,
\bea
M_n&=&(-1)^{n+1}\sum_{\sigma\in S_{n-3}}\sum_{\a\in S_{j-1}}\sum_{\b\in S_{n-j-2}}A_n(1,\sigma_{2,j},
\sigma_{j+1, n-2}, n-1,n){\cal S}[\a_{\sigma(2),\sigma(j)}|\sigma_{2,j}]_{p_1}\cr
&&\times \tilde {\cal S}[\sigma_{j+1,n-2}|\b_{\sigma(j+1),\sigma(n-2)},n]_{p_n}\tilde A_n(\a_{\sigma(2),\sigma(j)},
1,n-1,\b_{\sigma(j+1),\sigma(n-1)},n)\label{KLTbohr}
\eea
but with only $(n-3)!$ manifest symmetry.

{\bf 4-point}

The color factors for the SYM 4-point functions are $c(ij;kl)\equiv f^{a_ia_jb}f^{ba_ka_l}$, or in an obvious notation
\be
c_u\equiv \tilde f^{a_4a_2b}\tilde f^{ba_3a_1};\;\;\;
c_s\equiv \tilde f^{a_1a_2b}\tilde f^{ba_3a_4};\;\;\;
c_t\equiv \tilde f^{a_2a_3b}\tilde f^{ba_4a_1}
\ee
satisfying $c_u=c_s-c_t$ (thus with a slightly different sign convention than in the general case).
One can also find numerators $n_s,n_t,n_u$ satisfying the same Jacobi identity $n_u=n_s-n_t$. 
Using cyclicity and reflection invariance one is left with 3 amplitudes, which can be written in terms of the 
numerators as 
\bea
&&A_4(1234)=\frac{n_s}{s}+\frac{n_t}{t}\cr
&&A_4(1342)=-\frac{n_u}{u}-\frac{n_s}{s}\cr
&&A_4(1423)=-\frac{n_t}{t}+\frac{n_u}{u}
\eea
The color-dressed amplitude is then written as
\bea
{\cal A}_4^{tree}(1234)&=&g^2[\tilde f^{a_2a_3c}\tilde f^{ca_4a_1}A_4(1234)+\tilde f^{a_1a_3c}\tilde f^{ca_4a_2}A_4(2134)]\cr
&=&g^2\left[\frac{c_tn_t}{t}+\frac{c_un_u}{u}+\frac{c_sn_s}{s}\right]
\eea
and the gravity amplitude is 
\be
-iM_4(1234)=\frac{n_s^2}{s}+\frac{n_t^2}{t}+\frac{n_u^2}{u}
\ee
The KK basis is made up of 2 amplitudes, $A_4^{tree}(1234)$ and $A_4^{tree}(1324)$, but using the numerator gauge 
invariance, we can prove that only one is independent, i.e. $A_4^{tree}(1234)$, and we can isolate the helicity 
information in a factor $K$. 
The non-singular KLT relation is 
\be
M_{4}(1234)=-s_{12}A(1234)A(1243)\label{4pointklt}
\ee
and can be rewritten as the fact that the helicity factor of $M_4$ is the square of the helicity factor of 
$A_4$. 

{\bf 5-point}

The color factors are defined as
\be
c(ij;k;lm)\equiv \tilde f^{a_ia_jb}\tilde f^{ba_kc}\tilde f^{clm}
\ee
One can define corresponding numerators $n(ij;k;lm)$ satisfying the relations 
\bea
&&n(ij;k;lm)=-n(ji;k;lm)=-n(ij;k;ml)\cr
&&n(ij;k;lm)=-n(ml;k;ji)
\eea
besides the Jacobi identities
\be
n(ij;k;lm)+n(ki;j;lm)+n(jk;i;lm)=0
\ee
The color-dressed amplitude is 
\be
{\cal A}_5^{tree}=g^3\sum_{15\; terms}\frac{n(ij;k;lm)c(ij;k;lm)}{s_{ij}s_{lm}}
\ee

Then the gravity amplitude is 
\be
-iM_5(12345)=\sum_{i=1}^{15}\frac{(n_i)^2}{(s_{ab}s_{cd})_i}=\sum_{15\; terms} \frac{(n(ij;k;lm))^2}{s_{ij}s_{lm}}\label{schematic}
\ee

The usual KLT relation is 
\be
M_5=s_{12}s_{34}A(12345)\tilde A(21435)
+s_{13}s_{24}A(13245)\tilde A(31425)\label{usualKLT}
\ee
and has $(n-3)!=2!$ symmetry, whereas the KLT relations (\ref{KLTbohr}) become, explicitly,
\bea
M_5&=&\sum_{\sigma,\tilde \sigma\in S_2}\tilde A(45,\tilde\sigma_{23},1)A(1,\sigma_{23},45)S[\tilde \sigma_{2,3}|\sigma_{2,3}]_{p_1}\cr
&=&s_{12}s_{13}(A(45231)A(12345)+A(45321)A(13245))+s_{13}(s_{12}+s_{23})A(45231)A(13245)\cr
&&+s_{12}(s_{13}+s_{23})A(45321)A(12345)\cr
M_5&=&\sum_{\sigma,\tilde \sigma\in S_2}\tilde A(14,\tilde\sigma_{23},5)A(1,\sigma_{23},45)\tilde S[\sigma_{2,3}|\tilde \sigma_{2,3}]_{p_4}\cr
&=&s_{24}s_{34}[A(12345)A(14235)+A(13245)A(14325)]+
s_{34}(s_{24}+s_{23})A(12345)A(14325)\cr
&&+s_{24}(s_{34}+s_{23})A(13245)A(14235)\label{KLT5}
\eea 
and have $(n-3)!=2!$ symmetry. 

\section{1-loop expansion}

In this section we define the 1-loop expansion of amplitudes and write relations among the 1-loop amplitudes.

At 1-loop, there are several possible trace structures, unlike at tree level, specifically
\bea
{\cal A}_n^{1-loop}(12...n)&=&g^n \sum_{j=1}^{[n/2]+1}\sum_{\sigma\in S_n/S_{n;j}}G_{2n;j}(\sigma)A_{n;j}(\sigma(1)...\sigma(n))\cr
G_{rn;1}(1)&=&N_c\Tr(T^{a_1}...T^{a_n})\cr
G_{rn;j}(1)&=&\Tr(T^{a_1}...T^{a_{j-1}})\Tr(T^{a_j}...T^{a_n})\cr
A_{n;j}(12...,j-1,j,j+1,...n)&=&(-1)^{j-1}\sum_{\sigma\in COP\{\a\},\{\b\}}A_{n;1}(\sigma)
\eea
where $\a_i\in \{\a\}=\{j-1,j-2,...,2,1\};\b_i\in \{\b\}=\{j,j+1,...,n-1,n\}$ and $COP\{\a\},\{\b\}$ are permutations with the $n$ fixed and 
keeping $\{\a\}$ and $\{\b\}$ fixed up to cyclic permutations. Here $A_{n;j}(12...n)$ are 1-loop color-ordered 
amplitudes. 

But not all of the $A_{n;j}$ are independent. In fact, there are relations among them, which can be derived from 
group theory. 

In particular, for 5-point amplitudes, one has a single-trace amplitude $A_{5;1}$ and a double-trace
amplitude $A_{5;3}$ related by \cite{Bern:1991aq}
\be
A_{5;3}(45123)=\sum_{\sigma\in COP_4^{123}}A_{5;1}(\sigma(1),...,\sigma(4),5)\label{12trace}
\ee

The single-trace amplitude is given by
\be
A_5^{(1,0)}(12345)\equiv A_{5;1}(12345)=-\frac{1}{4}A_5(12345)\sum_{cyclic}F^{(1)}(s,t,m^2)
\ee
where
\be
F^{(1)}(s,t,m^2)=st I^{(1)}_5 (s,t,m^2)
\ee
is the dimensionless box, and $I^{(1)}(s,t,m^2)$ is the 1-loop scalar box integral with momenta 3,4 in the same corner
and $m^2=P^2=(p_3+p_4)^2$.

Substituting in (\ref{12trace}), we find
\be
A_{5;3}(fg;hij)=\sum_{abcde\in 30 \; fixed \; terms}F(cde;ab)[s_{abcde;+;fghij}A(abcde)+s_{abcde;-;fghij}A(abedc)]
\ee
Here $s_{abcde;\pm;fghij}$ are signs, defined as follows. The relative sign is plus if $ab$ belong to $hij$, 
and minus otherwise, and the overall sign is plus if the permutation of $hij$ inside $abcde$ is even, and minus 
if it is odd.

The 1-loop ${\cal N}=8$ supergravity amplitude is \cite{Bern:2006kd}
\bea
&&M_5^{1-loop}(12q_3q_2q_1)=-\frac{1}{2}\sum_{perms}s_{q_2q_1}s_{12}^2s_{2q_3}^2A(12q_3q_2q_1)A(12q_3q_1q_2)\times\cr
&&\times\int\frac{d^Dl_1}{
(2\pi)^D}\frac{1}{l_1^2l_2^2l_3^2l_4^2}+{\cal O}(\epsilon)\label{sugra2}
\eea

We can rewrite this in terms of the scalar 1m box $I(123,45)$ (with momenta 4,5 on the same corner of the box) and 
the dimensionless box $F(123;45)$, 
\bea
M_{5,sugra}^{1-loop}(12q_3q_2q_1)&=&-\frac{1}{2}\sum_{30\; perms}s_{q_2q_1}s_{12}s_{2q_3}A(12q_3q_2q_1)A(12q_3q_1q_2)[s_{12}s_{2q_3}I(12q_3;q_2q_1)]\cr
&=&-\frac{1}{2}\sum_{30\; perms}s_{q_2q_1}s_{12}s_{2q_3}A(12q_3q_2q_1)A(12q_3q_1q_2)F(12q_3;q_2q_1)
\eea
or
\be
M_{5,sugra}^{1-loop}(12345)=-\frac{1}{2}\sum_{30 perms}F(cde;ab)s_{cd}s_{de}s_{ab}A(cdeab)A(cdeba)\label{1loopsugra}
\ee

\section{IR behaviour and KLT relation}\label{irklt}

In this section we derive the main result of the paper, a KLT relation obtained by comparing 1-loop divergences in ${\cal N}=4$ SYM and 
${\cal N}=8$ supergravity. 
We will also use the information described here in section \ref{symsugrasection}, 
where we propose a relation between SYM and supergravity amplitudes, valid at least in the IR.

The IR behaviour of the 1-loop 1m scalar box is
\bea
I_{4,1m}(s,t,m^2)&=&\frac{r_\Gamma}{s_{12}s_{23}}\Big\{\frac{2}{\epsilon^2}[(-s_{12})^{-\epsilon}
+(-s_{23})^{-\epsilon}-(-s_{45})^{-\epsilon}]+{\rm finite}\Big\}\Rightarrow\cr
F(cde;ab)&\simeq& \frac{r_\Gamma}{\epsilon^2}[(-s_{cd})^{-\epsilon}+(-s_{de})^{-\epsilon}-(-s_{ab})^{-\epsilon}]
+{\rm finite}\cr
r_\Gamma&=&\frac{\Gamma(1+\epsilon)\Gamma^2(1-\epsilon)}{\Gamma(1-2\epsilon)},
\eea
where $D=4-2\epsilon$, 
giving for the {\bf IR behaviour of the double-trace 1-loop SYM amplitude} $A_{5;3}$
\bea
A_{5;3}(fg;hij)&=&\sum_{abcd\in 30\;{\rm terms}}F(cde;ab)[s_{abcde;+;fghij}A(abcde)+s_{abcde;-;fghij}A(abedc)]\cr
&\sim&\frac{r_\Gamma}{\epsilon^2}\sum_{abcd\in 30\;{\rm terms}}
[s_{cd}^{-\epsilon}+s_{de}^{-\epsilon}-s_{ab}^{-\epsilon}][s_{abcde;+;fghij}A(abcde)\cr
&&+s_{abcde;-;fghij}A(abedc)]
\eea
Organizing the coefficients of each divergence we find
\be
A_{5;3}(fg;lmn)\simeq \frac{r_\Gamma}{\epsilon^2}\sum_{i<j}(-s_{ij})^{-\epsilon}\sum_{abc\neq i,j}\epsilon_{lmn}[A(ijabc)]
\ee
where $\epsilon_{lmn}[A(ijabc)] $ means $A(ijabc)$ is multiplied by the sign of the permutation of $l,m,n$ inside 
$i,j,a,b,c$, and the sum over $a,b,c$ contains all the 6 terms of the arbitrary permutation of the $a,b,c\neq i,j$.

The leading ($1/\epsilon^2$) divergence of $A_{5;3}(45;123)$ is given by
\be
\sum_{i<j}\sum_{abc\neq i,j}\epsilon_{123}[A(ijabc)]\label{leadingIRgen}
\ee
which by explicit evaluation can be rewritten as 
\bea
&&5[A(12345)+A(12354)+A(12435)+A(12453)+A(12534)+A(12543)\cr
&&-A(13254)-A(13245)+A(14235)+A(14253)+A(15234)+A(15243)]\label{leadingIR}
\eea
The leading IR divergence of $A_{5;3}$ is $1/\epsilon$, as expected from a generalization of the subleading color amplitude of the 4-gluon 
amplitude \cite{Naculich:2008ew,Naculich:2009cv}.
Thus if there is no $1/\epsilon^2$ divergence, (\ref{leadingIR}) must be zero, a fact that we will prove later in the paper.

For {\bf ${\cal N}=8$ supergravity}, we obtain the IR behavior
\bea
M_{5,sugra}^{1-loop}(12345)&=&-\frac{1}{2}\sum_{30 perms}F(cde;ab)s_{cd}s_{de}s_{ab}A(cdeab)A(cdeba)\cr
&\simeq & -\frac{1}{2\epsilon^2}\sum_{30 perms}[s_{cd}^{-\epsilon}+s_{de}^{-\epsilon}-s_{ab}^{-\epsilon}]s_{cd}s_{de}s_{ab}A(cdeab)A(cdeba)
\eea
Organizing the terms by IR divergences, we obtain
\bea
&&M_{5,sugra}^{1-loop}\simeq \frac{1}{\epsilon^2}\sum_{i<j}s_{ij}^{1-\epsilon}
\times \Big[\sum_d s_{cd}s_{de}A(ijcde)A(ijedc)\cr
&&+\sum_c s_{ic}s_{ab}A(ijabc)A(ijbac)+\sum_c 
s_{jc}s_{ab}A(ijcba)A(ijcab)\Big]
\eea

On the other hand, we know that the IR behaviour of the 1-loop $n-$point supergravity amplitude is 
\cite{Dunbar:1995ed}
\be
M_n^{1-loop}(1...n)\simeq \frac{1}{\epsilon^2}M_{n}^{tree}(1...n)\sum_{i<j} s_{ij}^{1-\epsilon}
\ee
which means that we must have the KLT relation
\bea
M_5^{tree}(12345)&=&\sum_d s_{cd}s_{de}A(ijcde)A(ijedc)+\sum_c s_{ic}s_{ab}A(ijabc)A(ijbac)\cr
&&+\sum_c s_{jc}s_{ab}A(ijcba)A(ijcab),\hspace{3cm}
\forall (ij)\label{KLTus}
\eea
This formula is the main result of our paper. Note that it has the larger manifest symmetry of $2\times (n-2)!=2\times 3!$, which is larger 
than that of (\ref{KLTbohr}), and even 
larger than that of (\ref{KLTold}). Moreover (\ref{KLTus}) has no need to be regularized.

\section{Testing the new KLT relations}

\subsection{Explicit proof}

We have derived the tree-level KLT formula (\ref{KLTus}) using the IR behavior of 1-loop computations. However it is useful to prove it explicitly. 
To do so, we use the helicity spinor formalism and the Parke-Taylor formula \cite{Parke:1986gb}, which states that 
\be
A_n(1^+2^+...i^-...j^-...n^+)=\frac{\langle ij\rangle ^4}{\langle 12\rangle\langle23\rangle...\langle n1\rangle}
\ee
or for our case, for instance choosing $1^-2^-$,
\be
A_5(1^-2^-3^+4^+5^+)=\frac{\langle12\rangle^4}{\langle12\rangle\langle23\rangle\langle34\rangle\langle45\rangle\langle51\rangle}
\ee
A similar formula exists for the supergravity amplitude \cite{Bern:1998sv}
\be
M_5(1^-2^-3^+4^+5^+)=\frac{\langle 12\rangle^8\epsilon(1234)}{N(5)}
\ee
where 
\bea
\epsilon(ijkl)&=&4i\epsilon_{\mu\nu\rho\sigma}k_i^\mu k_j^\nu k_k^\rho k_l^\sigma\cr
N(5)&=&\prod_{i=1}^4\prod_{j=i+1}^5\langle ij\rangle
\eea

We want to prove a specific case of (\ref{KLTus}), namely
\bea
&&M_5(12345)=\cr
&=&s_{34}s_{45}A(12345)A(12543)+s_{53}s_{34}A(12534)A(12435)+s_{45}s_{53}A(12453)A(12354)\cr
&&+s_{23}s_{45}A(12345)A(12354)+s_{24}s_{35}A(12435)A(12453)+s_{25}s_{34}A(12534)A(12543)\cr
&&+s_{13}s_{45}A(21345)A(21354)+s_{14}s_{35}A(21435)A(21453)+s_{15}s_{34}A(21534)A(21543)\cr
&&\label{KLTnew}
\eea
The others follow from permutations and symmetry.

We make use of the helicity spinor identities
\bea
&&\langle ij\rangle=-\langle ji\rangle;\;\; [ij]=-[ji];\;\; s_{ij}=\langle ij\rangle[ji]\cr
&&\langle ij\rangle \langle kl\rangle =\langle ik\rangle\langle jl\rangle +\langle il\rangle\langle kj\rangle ;\;\;
[ij][kl]=[ik][jl]+[il][kj]\cr
&&\sum_{i=1,i\neq j,k}^n[ji]\langle ik\rangle =0\cr
&&\langle ij\rangle [jk]\langle kl\rangle[li]=\frac{1}{2}[s_{ij}s_{kl}-s_{ik}s_{jl}+s_{il}s_{jk}-\epsilon(ijkl)]\Rightarrow\cr
&&\epsilon(ijkl)=[ij]\langle jk\rangle[kl]\langle li\rangle -\langle ij\rangle[jk]\langle kl\rangle [li]\label{helisp}
\eea
Then we find that the right hand side of (\ref{KLTnew}) becomes
\bea
&&\frac{\langle 12\rangle^8}{\langle 12\rangle N(5)}\Big[[43][54]\langle 14\rangle\langle 24\rangle\langle 35\rangle+[35][43]\langle 13\rangle
\langle 23\rangle\langle54\rangle +[54][35]\langle 15\rangle\langle25\rangle\langle 43\rangle\cr
&& +[32][45]\langle 13\rangle\langle 25\rangle\langle 24\rangle 
+[42][53]\langle 14\rangle\langle23\rangle\langle 25\rangle +[52][34]\langle 15\rangle\langle 24\rangle\langle 23\rangle \cr
&&+[45][53]\langle 14\rangle\langle23\rangle\langle 51\rangle +[53][14]\langle 15\rangle\langle 24\rangle\langle 31\rangle 
+[34][15]\langle 13\rangle\langle25\rangle\langle 41\rangle \Big]\cr
&=&\sum_{9terms}\frac{\langle 12\rangle ^8[ij][kl]\langle 1i\rangle 
\langle 2l\rangle\langle jk\rangle }{\langle 12\rangle N(5)}\label{intermediate}
\eea
Note that the denominator of (\ref{intermediate}) is an even function under $1\leftrightarrow 2$, as is the numerator. 

To proceed, we make further use of the identities (\ref{helisp}), but in doing so we encounter the kinematic invariants $s_{ij}$, which 
are however not independent. At 5-points, there are 5 independent kinematic 
invariants, which can be taken to be $s_{i,i+1}$. Using $k_1+..+k_5=0$ and $k_i^2=0$, we can find the 
remaing $s_{ij}$'s in terms of them as 
\bea
s_{13}&=&s_{45}-s_{12}-s_{23}\cr
s_{24}&=& s_{51}-s_{23}-s_{34}\cr
s_{35}&=&s_{12}-s_{34}-s_{45}\cr
s_{41}&=&s_{23}-s_{45}-s_{51}\cr
s_{52}&=&s_{34}-s_{51}-s_{12}\label{kinematical}
\eea
Momentum conservation implies also that we have for instance $\epsilon(1235)=-\epsilon(1234)$, etc.
Finally, after a long calculation, we find that the right hand side of (\ref{KLTnew}) is equal to
\be
\frac{\langle 12\rangle ^8\epsilon(1234)}{N(5)}=M_5(1^-2^-3^+4^+5^+)
\ee
proving our relation.

We note however that this proof is not algorithmic, so it is unclear how we could generalize it to other cases, e.g. to higher $n$-point 
amplitudes.

\subsection{Observations on the basis of independent amplitudes and other methods}

It would be useful to present a different, algorithmic, proof, as that may be useful for generalization to $n>5$.
A necessary first step for such a proof would be 
to find an independent basis of tree-level amplitudes. Then, also having a basis of independent kinematic invariants $s_{i,i+1}$, one should 
express the right hand side of our KLT relation (\ref{KLTus}) 
in terms of it, do the same for right hand side of the usual KLT (\ref{usualKLT}), and compare.

At 5-points, according to the general discussion, there should be $(n-3)!=2$ independent amplitudes, once we 
take into account the numerator gauge invariances, out of the $(n-2)!=6$ amplitudes of the KK basis.

We can take two directions at this point. One could choose a basis of two numerators, or a basis of two amplitudes. 
For the basis of numerators, we can first solve the 9 Jacobi identities to reduce the 15 numerators to 6, the 
same number as the KK basis. For instance, we can choose them, same as \cite{Vaman:2010ez}, to be
 $n_1, n_6, n_9, n_{12}, n_{14}, n_{15}$.
Then the procedure chosen in \cite{Bern:2010yg} is to use an explicit helicity configuration 
($1^-2^+3^+4^+5^-$), in which case one has
\be
n_2=n_3=n_4=n_5=n_7=n_8=n_{11}=0;\;\;\;
n_{10}=-n_{13}=n_6-n_1
\ee
and from the KK basis above only 2 are independent
\be
n_1=n_{15}=n_{12}=-i\frac{<15>^3}{<23><34>}[12][45];\;\;\;\;
n_6=n_{14}=n_9=-i\frac{<15>^3}{<23><34>}[14[25]
\ee

We can substitute these numerators in the KK basis written as in \cite{Bern:2010yg} 
and use the KK relations (\ref{KK}) to then write down 
all the amplitudes in terms of $n_1, n_6$. Then we can immediately check that (\ref{leadingIR}) is indeed zero. 
This proves that the leading IR divergence of $A_{5;3}$ is $1/\epsilon$.

But despite this fact, the basis of two numerators is not the minimal one, since we have 
\be
\frac{n_1}{n_6}=-\frac{<51>[12]<24>[45]}{<51>[14]<42>[25]}=
\frac{-s_{51}s_{24}+s_{52}s_{14}-s_{54}s_{12}+\epsilon(5124)}{+s_{51}s_{24}
+s_{52}s_{14}-s_{54}s_{12}+\epsilon(5124)}\label{proport}
\ee
and since $\epsilon(ijkl)$ is a Lorentz invariant, it can be in principle expressed in terms of the independent invariants $s_{i,i+1}$, therefore 
$n_1$ is linear in $n_6$, using only $s_{i,i+1}$, meaning there is only one independent numerator.

In fact, we have checked that if we treat $n_1$ and $n_6$ as independent, 
we find that the previously proven KLT relations (\ref{KLT5}) seem not to be satisfied, meaning that we actually need to express everything 
in terms of a single independent numerator to obtain the expected result.

We could then use (\ref{proport}) to express everything in terms of a single independent numerator, and thus obtain an algorithmic method of 
checking relations, but unfortunately in practice it appears excessively  complicated.

An alternative, used in \cite{Vaman:2010ez}, is to write down a matrix constraint equation in the KK basis of 6 amplitudes, written formally as the 
vector $|A>$, of the type $C|A>=0$, but with $C$ of rank 4, meaning 
only 4 out of these 6 constraints are independent. We can then choose two apparently independent amplitudes, for instance $A_1\equiv A(12345)$
and $A_6\equiv A(13245)$, and express all the others terms of them. 

We have checked that if we do that, we get now an apparent mismatch for our KLT relation (\ref{KLTus}), which we have however already proven 
explicitly. That means that as seen before, in fact only one amplitude is truly independent, a fact that to our knowledge has not been appreciated
before. However we are not sure how to write the extra independent constraint on the vector $|A>$ and obtain an alternative algorithmic method of 
checking relations.

The existence of a single independent amplitude $A_5(12345)$ in turn means the possibility of writing a KLT relation of the type 
$M_5(12345)=B[A_5(12345)]^2$, with $B$ depending only on $s_{ij}$, 
exactly as in the case of 4-point amplitudes (\ref{4pointklt}), but we will not pursue this further. 

In fact, as we already described in section \ref{kltsection}, 
for general $n-$point amplitudes, on the $(n-2)!$ KK basis of amplitudes $|A>$, there is a linear matrix constraint (obtained from the generalized
gauge invariances acting on numerators)
$C|A>=0$, given explicitly in \cite{Vaman:2010ez}, where $C$ has rank $(n-3)(n-3)!$, equaling 4 in our $n=5$ case, and leaving $(n-3)!$ independent amplitudes,
2 in our case. Since we saw that for $n=5$ there is 
one more constraint, it means the general case must also be modified. An interesting possibility that should be explored is that in fact there is 
only a single independent amplitude for any $n$. That in turn would present
the possibility of a relation $M_n(12...n)=B[A_n(12...n)]^2$, with $B$ depending only on $s_{ij}$.

\section{Subleading SYM vs. supergravity relation}\label{symsugrasection}

%We will now use the information from section \ref{irklt} to find a relation between $A_{5;3}$ and $M_5^{1-loop}$ valid for the IR divergences, 
%and show how it could extend for the full amplitudes, similar to the relation at 4-points between 
%$A_{4;3}=A_4^{(1,1)}$ and $M_4^{1-loop}$ \cite{Naculich:2008ys}, $M_4^{1-loop}\sim K A_4^{1,1}$.

Let us recall the leading color, and subleading color-ordered, $L$-loop, 4-gluon amplitude
\bea
A_{4;1}&=&g^2a^L\Big[A_{4;1}^{(L,0)}+\frac{1}{N^2}A_{4;1}^{(L,2)}+...\Big]\cr
A_{4;3}&=&g^2 a^L\Big[\frac{1}{N}A_{4;3}^{(L,1)}+\frac{1}{N^3}A_{4;3}^{(L,3)}+...\Big]\cr
a&\equiv&\frac{g^2N}{8\pi^2}\left(4\pi e^{-\gamma}\right)^\epsilon
\eea
with the series ending at the $N$-independent amplitude $A^{(L,L)}$. The leading-color term $A^{(L,0)}$ comes from planar diagrams, while the 
subleading color amplitudes are $A^{(L,1)},...$, $A^{(L,L)}$. It was shown in \cite{Naculich:2008ew,Naculich:2009cv} that the leading IR behavior of 
the subleading color amplitude $A_{4;3}^{(L,k)}\sim 1/\epsilon^{2L-k}$, in contrast with the $1/\epsilon^{2L}$ IR behavior of $A_{4;1}^{(L,0)}$.
In particular $A_{4;3}^{(1,1)}\sim 1/\epsilon$. It was then shown \cite{Naculich:2008ew} that after the tree amplitudes are factorized off, the 
$L=1$ subleading color ${\cal N}=4$ SYM 4-point amplitude is proportional to the $L=1$, 4-graviton amplitude of ${\cal N}=8$ supergravity. 
That is, the 1-loop ${\cal N}=8$ supergravity amplitude can be expressed in terms of $A_{4;3}$, noting that they both have $1/\epsilon$ IR 
behavior. Similar results hold for $L=2$ as well.

Motivated by the fact that the leading IR divergence of the $n=5$-point  supergravity amplitude and of $A_{5;3}$ are both of order $1/\epsilon$
at 1-loop, in this section we investigate whether $M_5^{1-loop}$ can be expressed as a linear combination of $A_{5;3}$ amplitudes. We will 
use information from section \ref{irklt} and find a relation valid for IR divergences, and show how it could extend for the full amplitudes.

The strategy is as follows. Based on what happened at $4-$points at $1-$ and $2-$loops, we want to find 
$M_5^{1-loop}$ as a linear combination of the $A_{5;3}$ amplitudes. We will first analyze such a relation,
impose consistency conditions on it with the hope of obtaining a unique formula valid for IR divergences. We will then write down the formula 
that would be needed to be satisfied in order for our relation to extend to the full amplitudes.

In analogy with with the 4-point function, we would like to find a relation of the type
\bea
&&M^{1-loop}_{5,sugra}(12345)=\sum_{i\in fghij} \beta_i A_{5;3}(i)\cr
&&=\sum_{abcde\in 30\; fixed\; terms}F(cde;ab)
\sum_{i\in fghij}\beta_i[s_{abcde;+;i}A(abcde)+s_{abcde;-;i}
A(abedc)]\cr
&&\label{type}
\eea
On the other hand, we know that 
\be
M^{1-loop}_{5,sugra}(12345)=\sum_{abcde\in 30\; fixed \; terms}F(cde;ab)\a_{abcde}
\ee
where from (\ref{1loopsugra}), 
\be
\a_{abcde}=-\frac{1}{2}s_{cd}s_{de}s_{ab}A(cdeab)A(cdeba)
\ee
which means that we need
\be
\a_{abcde}=\sum_{i\in fghij}\beta_i[s_{abcde;+;i}A(abcde)+s_{abcde;-;i}
A(abedc)]\label{betaeq}
\ee
to be satisfied,
which are 30 equations for 10 unknowns ($\beta_i$), so (\ref{betaeq}) are not guaranteed to have solutions.

Let us understand the counting better. To start,
there are 20 independent $A_{5;3}$'s (terms in $(fghij)$), corresponding to $S_5/(Z_2\times Z_3)$, i.e. $5!/(2\times 3)=20$.
They can be found by making a permutation of $12345$. 
But we can reduce the amplitudes further. We note that the rule for the overall sign of the coefficient of the box is plus if the permutation of 
$hij$ inside $abcde$ is even and minus if it's odd means that
\be
A_{5;3}(fg;hij)=-A_{5;3}(fg;ihj)
\ee
and since we have also $Z_2\times Z_3$ symmetry, there are only 10 independent $A_{5;3}$'s, corresponding to picking $fg$ out of $1,2,3,4,5$. Then indeed, there are 10 unknowns $\beta_i$ as well, and we could also relabel 
$\b_i$ as $\b_{(fg)}$. As for the number of equations, 
there are as many as there are coefficients of boxes $F(cde;ab)$. Since the box has a symmetry under exchange 
of $a,b$ and of $c,e$, we have $5!/(2\times 2)=30$ terms, therefore 30 equations.

The 30 equations can then be rewritten, using the explicit form of $\a_{abcde}$, and a new notation that will 
prove useful, as
\be
-\frac{1}{2}s_{ab}s_{bc}s_{de}A(abcde)A(abced)=\sum_{fg;hij}\beta_{(fg)}\epsilon_{hij}[A(abcde)]\Big(1-\epsilon_{hij}(de)\frac{A(abced)}{A(abcde)}\Big)
\label{betaeqs}
\ee
where $\epsilon_{hij}(de)$ is plus if both $d,e$ belong to $h,i,j$, and minus otherwise.

In order to see if a unique solution for the $\b_i$ is possible, we will match the IR behaviours on the two sides of 
(\ref{type}). Expressing the IR behaviours of the lhs and the rhs, we get
\be
\frac{1}{\epsilon^2}M_5(12345)\sum_{i<j}s_{ij}(-s_{ij})^{-\epsilon}=\frac{r_\Gamma}{\epsilon^2}\sum_{k\in fg;lmn}\beta_k\sum_{i<j}
(-s_{ij})^{-\epsilon}\sum_{abc\neq i,j}\epsilon_{lmn}[A(ijabc)]]
\ee
which means that we need, using the vanishing of (\ref{leadingIR}),
\be
M_5(12345)s_{ij}=\sum_{k\in fg;lmn}\beta_k\sum_{abc\neq i,j}\epsilon_{lmn}[A(ijabc)]
\ee
If we denote the $M_5(12345)$ by just $M_5$, then the lhs is a vector column of $(ij)$, $M_5s_{ij}$. Also denote $\sum_{abc\neq i,j}\epsilon_{lmn}[A(ijabc)]$ as $N_{(ij),(fg)}$, so that
\be
N_{(ij),(fg)}\beta_{(fg)}=M_5 s_{ij}\Rightarrow [\beta_{(fg)}]=[N_{(ij),(fg)}]^{-1}M_5s_{ij}\label{coeff}
\ee
Note that the index $(fg)$ on the matrix $N$ has 10 values, and these values can also be identified by the $lmn$ of $\epsilon_{lmn}[A(ijabc)]$, since
it corresponds to the same 10 terms, picking out a group $(fg)$ or $(lmn)$ out of $1,2,3,4,5$. In $\epsilon_{lmn}[A(ijabc)]$, $l,m,n$ can be put in 
the order they appear in $1,2,3,4,5$. Since the notation is quite dense, let us give a few examples
\bea
N_{(12),(12)}&=&\sum_{perm.of(345)}\epsilon_{345}[A(12(345))]\cr
&=&A(12345)-A(12543)-A(12435)+A(12534)-A(12354)+A(12453)\cr
N_{(12),(13)}&=&\sum_{perm.of(345)}\epsilon_{245}[A(12(345))]\cr
&=&A(12345)-A(12543)+A(12435)-A(12534)-A(12354)+A(12453)\cr
N_{(13),(13)}&=&\sum_{perm.of(245)}\epsilon_{245}[A(13(245))]\cr
&=&A(13245)-A(13542)-A(13425)+A(13524)-A(13254)+A(13452)\cr
&&
\eea
At this point however we note that the vanishing of the leading IR divergence in (\ref{leadingIRgen}) means that 
\be
\sum_{(ij)}N_{(ij),(fg)}=0
\ee
i.e., that the matrix $N$ has rank 9 instead of 10. But, since $\sum_{(ij)}s_{ij}=0$ as we can check by summing over (\ref{kinematical}), 
in fact one of the 10 equations in (\ref{coeff}) is the sum of the other 9, and can be dropped. We need then to work with the corresponding 
$9\times 9$ reduced matrix $N_{red;(ij),(fg)}$ and give the 10th coefficient $\beta_{(fg)}$ an arbitrary value.

Therefore we have found a linear relation, (\ref{type}), with coefficients obtained from (\ref{coeff}), which is satisfied by the IR divergences, 
and containing an arbitrary parameter. 
Of course, it is still not clear that the remaining $\b_{(fg)}$ are unique. For that, one must calculate the rank of $N_{red}$. 
If its rank is less than 9, the solution is parametrized by more than 
one parameter, since then some of the remaining $\beta$'s will be undetermined. As the algebra is quite involved, 
we will leave this for further work.

In order to see if (\ref{type}) is satisfied also for the full amplitude, 
one must substitute the solution for $\b_{(fg)}$ back in (\ref{betaeqs}) and see if these equations are satisfied, since now we need to check 
whether the 30 equations are satisfied by substituting the 10 unknowns $\b_{(fg)}$ solved as in (\ref{coeff}). 
That is, we need to check whether 
\bea
&&-\frac{1}{2}s_{12}s_{23}s_{45}A(12345)A(12354)=\cr
&&\sum_{fg;hij}[N_{(ij),(fg)}]^{-1}M_5s_{ij}[s_{45123;+;(fg)}A(12345)-s_{45123;-;(fg)}A(12354)]
\eea
and 29 other equations. In the notation of (\ref{betaeqs}), we have
\bea
&&-\frac{1}{2}s_{ab}s_{bc}s_{de}A(abcde)A(abced)=\cr
&&=M_5\sum_{fg;lmn}[N_{(ij),(fg)}]^{-1}s_{ij}
\epsilon_{lmn}[A(abcde)]\Big(1-\epsilon_{hij}(de)\frac{A(abced)}{A(abcde)}\Big)\cr
&&N_{(ij),(fg)}\equiv \sum_{abc\neq i,j}\epsilon_{lmn}[A(ijabc)]\label{symsugra}
\eea
Here when we write $[N_{(ij),(fg)}]^{-1}s_{ij}$ we mean $[N_{red;(ij),(fg)}]^{-1}s_{ij}$ for 9 of the 10 values, and an arbitrary value for the 
tenth. Note that if these relations are true, they will also be of KLT type, since $N^{-1}\sim 1/A$, thus the relations are of 
the type $\alpha M_5= A A$, with $\a$ a coefficient containing $A/A $ terms. 
The relation looks however quite complicated, and we will leave it for further work.

\section{Comments on higher $n-$point amplitudes}

In principle the strategy used in this paper can be applied to higher $n-$point amplitudes by analogous methods. 
What is needed is an $n$-point generalization of the starting point; the relation between subleading color-ordered 
amplitudes and the leading color-ordered amplitude at 1-loop, which was eq. (\ref{12trace}), and a relation between the 
1-loop supergravity amplitude and the tree SYM amplitudes and scalar boxes, eq. (\ref{sugra2}). 

Then the same logic will apply. Namely one can analyze the IR behaviour of the results for ${\cal N}=4$ SYM and ${\cal N}=8$ supergravity 
at 1-loop, and compare these to the known behaviour, which would imply a relation among tree amplitudes from SYM, and a 
KLT-type relation from the supergravity. Finally, one can relate the subleading-color SYM and supergravity 
amplitudes, and use the consistency of the IR behaviour to fix the proposed relation.

We see that in order to complete this program, we need both the generalization of (\ref{12trace}) and of (\ref{sugra2}),
which are not yet available in the needed form, though it should be possible to obtain it. We hope to come back to this issue later. 

\section{Conclusions}

We have written a new kind of non-singular KLT relation for the $5-$point amplitudes, based on the analysis of 1-loop 
amplitudes. It has manifest $2(n-2)!$ symmetry, which is greater than previous known KLT relations, and also does not require regularization. 
We have also explicitly checked it using helicity spinor identities. We observed for SYM that at 5-points there is a single independent 
tree amplitude, but using this fact in an algorithmic method for checking identities turns out to be quite involved. 
Using these methods, we have proposed that the 5-point IR divergences of ${\cal N}=4$ SYM and ${\cal N}=8$ supergravity satisfy a linear 
relation: the ones of the 1-loop 5-point gravity amplitude are a linear combination of those of the subleading-color 1-loop SYM amplitudes 
$A_{5;3}$, in analogy with the known 4-point case \cite{Naculich:2008ew,Naculich:2009cv}. The consideration of this relation for the full amplitudes
needs further study. 

{\bf Acknowledgements} We are grateful to Lance Dixon for reading a preliminary draft of this paper, and encouraging us to complete an explicit 
proof of eq. (\ref{KLTus}).

The research of H.J. Schnitzer is supported in part by the DOE under grant DE-FG02-92ER40706.

\newpage

\bibliographystyle{utphys}
\bibliography{KLTpaper}

\Comment{
}

\end{document}